\documentstyle[11pt]{article}
\def\v1{\vspace{1cm}}
\def\be{\begin{equation}}
\def\ee{\end{equation}}
\def\bc{\begin{center}}
\def\ec{\end{center}}
\newcommand{\bea}{\begin{eqnarray}}
\newcommand{\eea}{\end{eqnarray}}

\begin{document}

\title{Topological invariant variables in QCD}
\author{
D. Blaschke, V. Pervushin\thanks{Joint Institute for Nuclear Research,
141980, Dubna, Russia.}, G. R\"opke\\
Fachbereich Physik, Universit\"{a}t Rostock, \\
D-18051 Rostock, Germany.}

\date{\today}

\maketitle

\medskip
%PACS number(s):04.60.-m, 04.20.Cv, 98.80.Hw (Quantum Gravity)
\medskip

%\newpage

\begin{abstract}
We show that the class of functions of topologically
nontrivial gauge transformations in QCD includes a zero-mode of
the Gauss law constraint.
The equivalent
unconstrained system compatible with Feynman's integral is derived
in terms of topological invariant variables,
where the zero-mode is identified with the winding number
collective variable and leads to the
dominance of the Wu-Yang monopole.
Physical consequences of  Feynman's path integral
in terms of the topological invariant variables are studied.
\end{abstract}
%%%%%%%%%%%%%%%%%%%%%%%%%%%%%%%%%%%%%%%%%%%%%%%%%%%%%%%%%%%%%%%%%%%%%%%%%
%%
%\documentstyle[12pt,psfig]{ioplppt}
%\newcommand{\be}{\begin{equation}}
%\newcommand{\ee}{\end{equation}}
%\newcommand{\bea}{\begin{eqnarray}}
%\newcommand{\eea}{\end{eqnarray}}
%\documentstyle{ioplppt}
%\begin{document}
%\jl{1}
%{\flushright{MPG-VT-UR 191/99}}\\[10mm]
%\title{Feynman Integral for the winding number variable in QCD}

%\author{D Blaschke\dag\ftnote{3}{To whom correspondence should be
%addressed.},
%V N Pervushin\ddag~ and G R\"opke\dag}

%\address{\dag\ Fachbereich Physik, Universit\"{a}t Rostock, D-18051
%Rostock,
%Germany}

%\address{\ddag\ Bogoliubov Laboratory of Theoretical Physics, Joint
%Institute
%for Nuclear Research, 141980 Dubna, Russia}

%\begin{abstract}
%\end{abstract}

%\pacs{11.15.-q, 12.38.Aw}
%\maketitle
%%%%%%%%%%%%%%%%%%%%%%%%%%%%%%%%%%%%%%%%%%%%%%%%%%%%%%%%%%%%%%%%%%%%%%%
\section{Introduction}

The anomalous decay processes in both electrodynamics
and chromodynamics  are described by  the spatial  integral
of the product of the magnetic and electric field strength tensors.
In non-Abelian field theory, this integral is known as the
time derivative of the winding number functional.

One of the main differences between QCD and QED is the fact of the
non-invariance of the QCD winding number functional with respect to
gauge transformations with stationary matrices (considered as maps of
the coordinate space into the color group one \cite{jac}).
This is a result of the nontrivial topological  structure of
the group of stationary gauge transformations.
The condition that the functional for the degree of the map is a finite
number (normalization) determines the 'winding number' class of
functions on which the gauge transformations act.
These functions decrease at spatial infinity as {\cal O}$(1/r)$.
This  means that
the dynamic gluon fields behave similar to the fields of point charges
(monopoles).

This difference between the classes of functions for dynamic fields
in QED and QCD is the principal peculiarity of the gauge theory
of strong interactions.
The present paper is devoted to a discussion of the dynamical consequences of
the 'winding number' class of functions in QCD.

The present approach is based on the fact that the winding number class of
functions contains the zero-mode of the  Gauss law constraint \cite{vp1,n}.
This zero-mode is identified with
the winding number as a collective variable~\cite{1,3}.

We derive an equivalent unconstrained system
compatible with the simplest canonical quantization scheme in the
form of Feynman's path integral.
From this path integral representation of the generating functional a
unitary perturbation theory for non-Abelian gauge theories can be obtained
(see ~\cite{f,al,fs}) which is similar intuitive Faddeev-Popov (FP)
scheme~\cite{fp}.

When comparing the scheme which will be developed in this paper with the
instanton approach~\cite{ins} the following basic differences are observed:
(i) the approach is formulated in the Minkowski space instead of the Euclidean
one, and (ii) the perturbation theory is formulated with physical states instead
of the classical vacuum ones.

The paper is organized as follows.
In Section 2 we discuss the statement of the problem in detail.
Section 3 is devoted to the derivation of the equivalent
unconstrained system with the zero-mode in the Yang-Mills theory.
In Section 4,
the generating functional for Green functions in the
form of the Feynman integral is constructed for QCD
as the basis for a solution of the problems of
the hadronization and confinement.

\section{The winding number class of functions}

We consider the winding number functional
\be \label{e1}
 X[A]=-\frac {1}{8\pi^2}\int\limits_V d^3x
\epsilon^{ijk}Tr \left[{\hat A}_i\partial_j{\hat A}_k -
 \frac 2 3 {\hat A}_i{\hat A}_j{\hat A}_k\right]
\ee
for the gluon gauge fields ${\hat A_\mu}=g\frac{\lambda^a }{2i} A_\mu^a~$,
in the non-Abelian $SU_c(3)$ theory with the action functional
\be \label{u}
W=\int d^4x
\left\{\frac{1}{2}({G^a_{0i}}^2- {B_i^a}^2)
+ \bar\psi[i\gamma^\mu(\partial _\mu+{\hat
A_\mu})
-m]\psi\right\}~,
\ee
where $\psi$ and $\bar \psi$ are the fermionic quark fields.
We use the conventional notations for the non-Abelian electric
and magnetic fields
\be \label{v}
G_{0i}^a = \partial_0 A^a_i - D_i^{ab}(A)A_0^b~,~~~~~~
B_i^a=\epsilon_{ijk}\left(\partial_jA_k^a+
\frac g 2f^{abc}A^b_jA_k^c\right)~,
\ee
as well as the covariant derivative
$D^{ab}_i(A):=\delta^{ab}\partial_i + gf^{acb} A_i^c$.

The action (\ref{u}) is invariant with respect to gauge transformations
$u(t,\vec x)$
\be \label{gauge1}
{\hat A}_{i}^u := u(t,\vec x)\left({\hat A}_{i} + \partial_i
\right)u^{-1}(t,\vec x),~~~~~~
\psi^u := u(t,\vec x)\psi~.
\ee
It is well-known \cite{fs} that the fixation of the gauge in the classical
equations of motion leaves the ambiguity in the choice of
initial data for the gauge fields due to the invariance
of the action with respect to the gauge transformations~(\ref{gauge1})
with stationary matrices $u(t,\vec{x})=v(\vec{x})$.

The group of the stationary gauge transformations $v(\vec{x})$ in the
coordinate space is topologically nontrivial and represents
the group of three-dimensional paths lying  on the three-dimensional
space of the $SU_c(3)$-manifold with the homotopy group
$\pi_{(3)}(SU_c(3))=Z$.
The whole group of the stationary gauge transformations is split into
topological classes marked by the integer number $n$ (the degree of the map)
which counts how many times a three-dimensional path turns around the
$SU(3)$-manifold when the coordinate $x_i$ runs over the space where it is
defined.
The stationary transformations $v^n(\vec{x})$ with $n=0$ are called the small
ones; and those with $n \neq 0$
\be \label{gnl}
{\hat A}_i^{(n)}:=v^{(n)}(\vec{x}){\hat A}_i(\vec{x})
{v^{(n)}(\vec{x})}^{-1}
+L^n_i~,~~~~L^n_i=v^{(n)}(\vec{x})\partial_i{v^{(n)}(\vec{x})}^{-1}
\ee
the large ones.

In QCD, the  winding number functional (\ref{e1})
is not invariant with respect to large gauge
transformations (\ref{gnl}) \cite{jac}
\be \label{gx}
X[A^{(n)}]=X[A]+{\cal N}_1[A,n]+{\cal N}_2[n]~,
\ee
where
\be \label{gn}\vspace{0.5cm}
{\cal N}_1[A,n]
=\frac {1}{8\pi^2}\int d^3x ~\epsilon^{ijk}~ Tr[\partial_i({\hat A}_j
L^n_k)]~,
\ee
\be \label{gn2}
{\cal N}_2[n]
=-\frac {1}{24\pi^2}\int d^3x ~\epsilon^{ijk}~ Tr[L^n_iL^n_jL^n_k]=n~.
\ee
It determines the degree of a map ${\cal N}_2[n]$ (see Eqs. (3.33), (3.36)
in \cite{jac1}).
The degree of a map ${\cal N}_2[n]=n$ as the condition of the
normalization  means that the large transformations
are given in the  class of functions with the spatial asymptotics
{\cal O}$(1/r)$.
Such a function $L^n_i$~(\ref{gnl}) is given by
\be \label{class0}
v^{(n)}(\vec{x})=\exp(n \hat \Phi(\vec{x})),~~~~~
\hat \Phi=- i \pi\frac{\lambda_A^a x^a}{r} f_0(r)~,
\ee
where the antisymmetric SU(3) matrices are denoted as
$$\lambda_A^1:=\lambda^2,~\lambda_A^2:=\lambda^5,~\lambda_A^3:=\lambda^7~,$$
and $r=|\vec x|$.
The function $f_0(r)$ satisfies the boundary conditions
\be \label{bcf0}
f_0(0)=0,~~~~~~~~~~~~~~
f_0(\infty)=1~,
\ee
so that the functions $L_i^n$ disappear at spatial infinity
$\sim$ {\cal O}$(1/r)$  but
can have nonvanishing surface integrals in Eq. (\ref{gx}).
We call the class of functions (\ref{class0}) the 'winding number' class of
functions.

The present paper is based on the evident fact that the dynamical field $A_i$
and its transformations $L^n_i$
both belong to the winding number class of functions.

The second fact is that the winding number class of functions includes
the topological excitations of the gluon system as a whole in the form of
the zero-mode $({\cal Z})$ of the non-Abelian Gauss law constraint
\cite{vp1,n}
\be \label{gc}
\frac{\delta W}{\delta A_0}=0~~~~~ \Rightarrow~~~~ D_{i}^{ab}(A)G_{0i}^b= - j_0^a ,
\ee
where $j_\mu^a=g\bar \psi \frac{\lambda^a}{2} \gamma_\mu\psi$
is the quark current.
The Gauss law takes the form of
an inhomogeneous equation for the time-like component $A_0$
of the gauge field
\be\label{gaussd}
(D^2(A))^{ac} { A_0}^c = D_i^{ac}(A)\partial_0 A_i^c+ j_0^a~.
\ee
A general solution of this inhomogeneous equation
is a sum of the solution ${\cal Z}^a$  of the homogeneous equation
\be\label{zm}
(D^2(A))^{ab}{\cal Z}^b=0~,
\ee
i.e., a zero mode of the Gauss law constraint, and a particular solution
${\tilde A}_0^a$ of the inhomogeneous one~(\ref{gaussd})
\be \label{genl}
A_0^a = {\cal Z}^a + {\tilde A}^a_0~.
\ee

It is the central point of our paper, that  the
zero-mode ${\cal Z}^a$  at the spatial infinity can be represented
in the form of the product of a new topological variable $ \dot N(t)$
and a phase $\Phi_0(\vec{x})$
\be \label{ass}
\hat {\cal Z}(t,\vec{x})|_{\rm asymptotics}=\dot N(t)\hat \Phi_0(\vec{x})~,
\ee
where the phase $\Phi_0(\vec{x})$ belongs to the winding number class of
functions (\ref{class0})
\be \label{as0}
\hat \Phi_0=- i \pi\frac{\lambda_A^a x^a}{r} f_0(r)~,~~~~~~~~~~~
f_0(0)=0,~~~~~~~~~~~~~~
f_0(\infty)=1~.
\ee
It is determined in the asymptotical region by the equation
\be\label{zm2}
(D^2)^{ab}({\Phi}_i){\Phi}_0^b(\vec{x})=0~,
\ee
where $\Phi_i(\vec{x})$ the asymptotics of the dynamical gluon fields $A_i$
given in the same class of functions
\be \label{ass1}
\hat A_i(t,\vec{x})|_{\rm asymptotics} = \hat {\Phi}_i(\vec{x})
= - i \frac{\lambda_A^a}{2}\epsilon_{iak}\frac{x^k}{r^2}f_1(r)~,
\ee
with the boundary conditions
\be \label{bcf1}
f_1(0)=0,~~~~~~~~~~~~~~~f_1(\infty)=1.
\ee
In this case, the single one-parametric variable $N(t)$
reproduces the topological degeneracy
of all field variables, provided
the separation of the zero-mode phase factors
of the topological degeneracy from the topological invariant
variables $A^I,\psi^I$ of perturbation theory
(i.e., the variables without topological degeneracy).
This separation is fulfilled by the gauge transformations
\be \label{gt1}
0=U_{\cal Z}(\hat {\cal Z}+\partial_0)U_{\cal Z}^{-1}~,
\ee
$$
{\hat A}_i=U_{\cal Z}({\hat A}^I+\partial_i)U_{\cal Z}^{-1},~~~
\psi=U_{\cal Z}\psi^I~,
$$
where the spatial asymptotics of $U_{\cal Z}$ is
\be \label{UZ}
U_{\cal Z}=T\exp[\int\limits^{t} dt'
\hat {\cal Z}(t',\vec{x})]|_{\rm asymptotics}
=\exp[N(t)\hat \Phi_0(\vec{x})]~.
\ee
A known example of fields $(\Phi_i,\Phi_0)$ which satisfy Eq. (\ref{zm2})
is the Bogomol'ny-Prasad-Sommerfield (BPS) monopole $(\Phi_i^{BPS})$
with the Higgs field $(\Phi_0^{BPS})$ \cite{bps}
\begin{eqnarray} \label{bps}
\Phi_i^{BPS}=
- i \frac{\lambda_A^a}{2}\epsilon_{iak}\frac{x^k}{r^2}f_1^{BPS}(r)~,~~~~~~~~
f_1^{BPS}(r)&=& 1 - \frac{r}{\epsilon \sinh(r/\epsilon)}~,\\
\hat \Phi_0^{BPS}=- i \pi\frac{\lambda_A^a x^a}{r} f_0^{BPS}(r)~,~~~~~~~~~~~
f_0^{BPS}(r)&=& \frac{1}{\tanh(r/\epsilon)}-\frac{\epsilon}{r}~,
\end{eqnarray}
in the limit $\epsilon \to 0$, where $\epsilon$ is the size of the monopole.
For $\epsilon \to 0$ the BPS monopole goes over into the
Wu-Yang monopole \cite{wy} without singularity at the origin
(as the solution of classical equations).

In this case, the winding number
functional (\ref{e1})  takes the form of the sum of the invariant term
and the new topological variable $N(t)$
\be \label{gxN}
X[A]=X[A^I]+{\cal N}_1[{\Phi}_i,N(t)]+{\cal N}_2[N(t)]=X[A^I]+N(t)~.
\ee
This can be seen ~\cite{jac1} from the fact that
for  noninteger $n=N(t)$ the responses of the winding number (\ref{e1})
are equal to
$${\cal N}_1[{\Phi }_i,N(t)]={\sin[2\pi N(t)]}/{2\pi},
$$
and
$${\cal N}_2[N(t)]= \{N(t)-{\sin[2\pi N(t)]}/{2\pi}\}~,
$$
so that the sinus terms cancel each other in the sum (\ref{gxN}).
The function $N(t)$ represents a one-dimensional parametrization of the
transition between different maps of the homotopy group.

The topological degeneration of initial data
means that the points $N(t)$ and $N(t)-n$ are physically equivalent.
Thus, the configuration space of the physical variables of QCD
has the topology of a cylinder where the role of the degree of freedom
which rotates around the cylinder is played by the topological
variable $N(t)$ as the zero mode of the first class constraint.
This zero mode can be revealed by only the explicit resolving the Gauss law
constraint~\cite{n}.
In the contrast to the zero mode of the
second class constraint (i.e., gauge) treated as the Gribov
ambiguity~\cite{g}, the zero mode of the first
class constraint is the inexorable consequence of internal dynamics,
which can be implicitly lost by the standard gauge-fixing scheme.

The present paper is devoted to the derivation of an equivalent unconstrained
system with the topological variable $N(t)$  and to the construction of
the generating functional for a unitary perturbation theory.

\section{Yang-Mills theory}

\subsection{Constraining}

Recall that for a long time the problem of quantization of non-Abelian
constrained systems was considered as the greatest challenge of
relativistic quantum field theory~\cite{sch}. The success in the description
of the quantum dynamics of such systems was the unitary perturbation theory in
the form of the Faddeev-Popov functional integral~\cite{fp} used for the
proof of the renormalizability of the Standard Model (Nobel Prize
of 1999 for 't Hooft and Veltman).
However, the foundation of the intuitive Faddeev-Popov integral
for the Yang-Mills (YM) theory was achieved by Faddeev~\cite{f} on the way of
the construction of an 'equivalent unconstrained system' compatible
with the simplest canonical scheme of quantization in the form
of the Feynman integral over independent physical variables
\be\label{ym}
 Z_{YM}^*[J]=\int \prod\limits_{t,x}\left\{
\prod\limits_{a=1}^{3}
\frac{[d^2 A_a^T d^2 E_a^T]}{2\pi}\right\}
\exp \left\{i W^*_{YM} + i\int d^4x J^a_i {A^a_i}^T\right\}~.
\ee
The action of this equivalent unconstrained system
\be \label{ymu}
W^*_{YM} =\int d^4x \left\{{E_k^a}^T \cdot {\dot A}_k^{aT}-
\frac{1}{2}\left\{ ({E_k^a}^T)^2
+B_k^2(A^T)+(\partial_k \sigma^a)^2\right\}
\right\}~,
\ee
where $\sigma^a$ satisfies the equation
$$
D(A)^{bc}_i\partial_i \sigma^c= -g\epsilon^{abc} A^a_i {E^c_i}^T~,
$$
is derived by the {\bf constraining}
\be \label{ymc}
W^*_{YM}=W_{YM}(\rm{constraint})
\ee
the initial YM action in the first order formalism
\be \label{ymfo}
W_{YM}=\int d^4x \left\{F_{0i}^a G_{0i}^a -
\frac{1}{2}[F_{0i}^2+B_i^2] \right\}~~~~~~~~~~~~
\left( G_{0i}^a= \dot A_i^a-D^{ab}_i(A)A_0^b\right)~.
\ee
The {\bf constraining}~(\ref{ymc}) means the substitution of the explicit
solution of the Gauss law constraint (see in details~\cite{al})
\be  \label{gcym}
D^{ab}_i(A) F^b_i=0
\ee
obtained by decomposing the electrical components of
the field strength tensor $F_{0i}$ into transverse ${F^a_{0i}}^T={E^a_i}^T$
and longitudinal ${F^a_{0i}}^L=-\partial_i \sigma^a$ parts
\be \label{longb1}
F^a_{0i}={E^a_i}^T -\partial_i \sigma^a;~~~
\partial_k  {E^b_k}^T =0~.
\ee
This decomposition is compatible with the perturbative gauge
\be \label{gym}
\partial_i A^a_i:=\partial_i {A^a_i}^T=0~.
\ee

\subsection{Constraining with the zero mode}

To formulate an equivalent unconstrained system for the YM theory
in the winding number class of functions in the presence
the zero mode ${\cal Z}^b$ of the Gauss law constraint~(\ref{gcym})
\be \label{edg1}
A_0^a={\cal Z}^a+\tilde A_0^a;~~~~~~
F_{0k}^a = - D_k^{ab}(A) {\cal Z}^b + {\tilde F_{0k}^a}~~~~~~
(~(D^2(A))^{ab}{\cal Z}^b=0~)~,
\ee
we shall follow two principles of the Faddeev foundation~\cite{f}:
i) the constraint-shell action~(\ref{ymc})
\be \label{cs1}
W_{YM}({\rm constraint}) =  {\cal W}_{YM}[{\cal Z}] + \tilde W_{YM}[\tilde F]~,
\ee
and ii) the choice of the gauge
\be \label{g1}
D_i^{ab}(\Phi_i)\tilde A_i^b:=0
\ee
compatible
with the perturbation theory around the SU(2) monopole $\Phi_i^a$
\be \label{evid}
\hat {\tilde F}_{0k}=U_{\cal Z} \hat F^I_{0k}U_{\cal Z}^{-1}~,~~~~
\hat A_i = U_{\cal Z}(\hat A_i^I+\partial_i )U_{\cal Z}^{-1}~,~~~
\hat A_i^I(t,\vec x)=\hat \Phi_i(\vec x) + \hat {\tilde A}_i(t,\vec x)~.
\ee
To present the action in the form~(\ref{cs1}), we can use
the evident decomposition
\be
 F^2=(-D{\cal Z}+\tilde F)^2=
 (D{\cal Z})^2- 2\tilde F D{\cal Z} + (\tilde F)^2=
 \partial ({\cal Z}(D{\cal Z}))- 2\partial ({\cal Z}\tilde F)+(\tilde F)^2
\ee
and the Gauss Eqs. $D\tilde F=0$ and $D^2{\cal Z}=0$ which show
that the zero mode part ${\cal W}_{YM}$ of the constraint-shell action~(\ref{cs1})
is the sum of two surface integrals
\be \label{gp}
{\cal W}_{YM}[{\cal Z}]=\int dt \int d^3x[\frac{1}{2}{\partial}_i({\cal Z}^a D_i^{ab}(A)
{\cal Z}^b)- {\partial}_i(\tilde F_{0i}^a{\cal Z}^a)]
={\cal W}^0 + {\cal W}^{\prime}~,
\ee
where the first one ${\cal W}^0$ is the kinetic term and the second one
${\cal W}^{\prime}$ describes the coupling of the zero-mode to the local
excitations.
These surface terms are determined by the asymptotics
of the fields $({\cal Z}^a, {A}_i^a)$ at spatial infinity (\ref{ass}),
(\ref{ass1})
which we denoted, in the SU(3) case,
by $(\dot N(t)\Phi_0^a(\vec{x}), {\Phi}_i^a(\vec{x}))$.
The fluctuations $\tilde F_{0i}^a$
belongs to the class of multipoles
and since the surface integral over monopole-multipole couplings vanishes, the
fluctuation part of the second term obviously drops out.
The substitution of the solution with the asymptotics (\ref{ass})
into the first surface term
Eq.~(\ref{gp}) leads to the zero-mode action
\begin{equation} \label{ktg}
{\cal W}^0=\frac{{\cal I}}{2}\int dt {\dot N(t)}^2 ~,
\ee
where
\be
{\cal I}=\int\limits
d^3x {\partial}_i[\Phi^a_0(\vec{x}) {\partial}_i \Phi^a_0(\vec{x})]=
(\frac{2\pi}{g})^2 {4 \pi}\epsilon~,
\ee
and
\be \label{C}
\epsilon =\lim\limits_{r_v \to \infty}{r_v^2f'_0(r_v)}~.
\end{equation}
$r_v$ is a value of $r$ at a boundary of a finite spatial volume.

The action for the equivalent unconstrained
system of the local excitations,
\be \label{lym}
\tilde W_{YM}[\tilde F]=\int d^4x \left\{E_k^a\cdot \dot{A}_k^{aI} -
\frac{1}{2}\left\{E_k^{2}
+B_k^2(A^I)+[D_k^{ab}({\Phi})\tilde{\sigma}^b]^2\right\}
\right\}~,
\ee
is obtained by decomposing the electrical components of
the field strength tensor $F_{0i}^I$ into transverse ${F^a_{0i}}^T=E^a_i$ and
longitudinal
${F^a_{0i}}^L=-D_i^{ab}({\Phi})\tilde \sigma^b$ parts, so that
\be \label{longb2}
F^{aI}_{0i}=E_i -D_i^{ab}({\Phi})\tilde \sigma^b;~~~
D_k^{ab}({\Phi}) E^b_k =0 ~.
\ee
Here the function $\tilde \sigma^b$ is determined from the Gauss equation
(\ref{gcym})
\begin{equation}\label{gaub1}
\left((D^2({\Phi}))^{ab} +
g\epsilon^{adc}{\tilde A}^{d}_i D_i^{cb}({\Phi})  \right) \tilde \sigma^b
 = - g\epsilon^{abc}{\tilde A}_i^{a} E_i^{c}~.
\end{equation}

\subsection{Feynman path integral}

The Feynman path integral over the independent variables
includes the integration over the topological variable $N(t)$
\be \label{fizm}
Z^{\rm Feynman}_{YM}[J]=\int \prod\limits_{t}dN(t)
\tilde Z_{YM}[J^U]~,
\ee
where
\be\label{schb}
\tilde Z[J^U]=\int \prod\limits_{t,x}\left\{
\prod\limits_{a=1}^{3}
\frac{[d^2 A_a^I d^2 E_a]}{2\pi}\right\}
\exp \left\{i{\cal W}_{YM} + i\tilde W_{YM} +i S[J^U] \right\}~.
\ee
As we have seen above, the functionals ${\tilde W},S$
are given in terms of the variables which
contain the nonperturbative phase factors $U=U_{\cal Z~}$ (\ref{UZ}) of the
topological degeneration of initial data. These factors disappear
in the  action $\tilde W$, but not in  the source
\be \label{sym}
S[J^U]=\int d^4x J^a_i A^a_i,~~~\hat A_i=U(\hat A_a^I +\partial_i)U^{-1}
\ee
what reflects the fact of the topological degeneration of the physical fields.

Thus, instead of the instanton
averaging over 'interpolations between different vacua',
in order to remove the topological degeneracy,
all Green functions should be averaged over values of the
topological variable and all possible angles of orientation of
the monopole unit vector
($\vec n=\vec x/r$)~(\ref{ass1}) in the group space.
This averaging leads to complete destructive
interference of colored Green functions and corresponding color amplitudes.
The complete destructive interference
of color amplitudes can be interpreted as confinement
in the spirit of the Feynman quark-parton duality \cite{vp1,vp3}.

\section{Quantum chromodynamics}

\subsection{Feynmann path integral}

QCD differs from the YM theory by the group SU(3) and quark fields.

The decomposition of the spatial components of the gluon fields
$A_i^a={\Phi}_i^a+\tilde A_i^a$ into the static
monopole ${\Phi}_i^a$
and the fluctuations can be used to define a perturbation theory with respect
to the dynamical gluon fields $\tilde A_i^a$.
The adequate gauge for this perturbation theory
is the covariant Coulomb gauge defined by
\be \label{ccg}
D^{ac}_i({\Phi}_i^b)\tilde A^{c}_i =0~.
\ee

After the separation of the phase factors of the topological degeneracy,
we obtain the similar path integral in terms of perturbative
fields ($E, A^I,\bar\psi^I,\psi^I$)
\be \label{fizm1}
Z_{QCD}^{\rm Feynman}[J,\eta,\bar\eta]=\int \prod\limits_{t}dN(t)
e^{i{\cal W}_0}
\tilde Z[J^U, \eta^U, \bar\eta^U] ,
\ee
where
\be\label{schb1}
\tilde Z[J^U, \eta^U, \bar\eta^UJ]=\int \prod\limits_{t,x}\left\{
[d \bar \psi^I d \psi^I]
\prod\limits_{a=1}^{8}
\frac{[d^2 A_a^I d^2 E_a]}{2\pi}\right\}
\exp \left\{i {\cal W}' + i \tilde W +
i S[J^U, \eta^U, \bar\eta^U]\right\}~,
\ee
\be
 \eta^U=U_{\cal Z}\eta~,
\ee
the zero-mode actions ${\cal W}_0$ and
${\cal W}'$
are defined by the expressions similar to~(\ref{gp})
and~(\ref{ktg}), in the first order formalism obtained by the Legandre
transformation
$$
\frac{1}{2}G_{0i}^2=G_{0i}F_{0i}-\frac{1}{2}F_{0i}^2.
$$
The quark part of ${\cal W}^{\prime}$ will be discussed later.
The action
for the equivalent unconstrained
system of the local excitations
\be\label{wl2}
\tilde  W =\int d^4x \left\{E_k^a\cdot \dot{A}_k^{aI} -
\frac{1}{2}\left\{E_k^{2}
+B_k^2(A^I)+[D_k^{ab}({\Phi})\tilde{\sigma}^b]^2\right\}
+ \bar\psi^I [i \gamma_{\mu}\partial^{\mu} - m]\psi^I
+j_i^a A_i^{aI} \right\}~,
\ee
is obtained by decomposing the electrical components of
the field strength tensor $F_{0i}^I$ (similarly~(\ref{evid}))
into transverse ${F^a_{0i}}^T=E^a_i$ and
longitudinal
${F^a_{0i}}^L=-D_i^{ab}({\Phi})\tilde \sigma^b$ parts, so that
\be \label{longb}
F^{aI}_{0i}=E_i -D_i^{ab}({\Phi})\tilde \sigma^b;~~~
D_k^{ab}({\Phi}) E^b_k =0 ~.
\ee
The function $\tilde \sigma^b$ is determined from the Gauss equation
(\ref{gc})
\begin{equation}\label{gaub}
\left((D^2({\Phi}))^{ab} +
gf^{adc}{\tilde A}^{d}_i D_i^{cb}({\Phi})  \right) \tilde \sigma^b
 = - j^b_{{\rm tot},0};
~~~~j^b_{{\rm tot},0}=gf^{abc}{\tilde A}_i^{a} E_i^{c}+j^b_0~.
\end{equation}
Equation (\ref{gaub}) can be formally solved by introducing a Green
function $G^{ab}(\vec{x},\vec{y})$ defined as the solution to
\begin{equation}\label{gaub0}
(D^2({\Phi}))^{ab}G^{bc}(\vec{x},\vec{y})=\delta^{ac}\delta^3(\vec{x}
-\vec{y})~.
\end{equation}
Then ${\tilde \sigma}^a$ can be expanded in a power series in $g$
\be \label{fse}
\tilde \sigma^{b}(t,\vec{x})=
- \int d^3yG^{bc}(\vec{x},\vec{y})j_{\rm tot,0}^c(t,\vec{y}) -
\int d^3yd^3zG^{bc}(\vec{x},\vec{y})gf^{cde}{\tilde A}_k^{d}(t,\vec{y})
G^{ef}(\vec{y},\vec{z})j_{\rm tot,0}^{f}(t,\vec{z}) - ...
\ee
in analogy to the standard perturbation theory \cite{al}.
The calculation of the Green function $G^{ab}(\vec{x},\vec{y})$ is given
in the Appendix.

\subsection{Dominance of the Wu-Yang monopole}

%Recall, that quantum field theory is formulated in the finite space-time
%to determine finite cross-section and energy density in the units of
%space volume and time.

The action ${\cal W}^0$ in the path integral~(\ref{schb1})
(given by~(\ref{ktg}))
describes a free 'rotator' with the momentum spectrum \cite{vp1}
\be \label{mom}
P_0= {\dot N}{M_N}= (2\pi k + \theta)~, ~~~~k=0,1,2,... ,
\ee
which follows from the constraint on the wave function for
physically equivalent points $N$ and $N+1$
$$\Psi(N+1)=\exp(i\theta)\Psi(N).$$

The action for the 'rotator' (\ref{ktg}) %$\int dt P^2/2{\cal I}$
compensates the action for the monopole background
$$
W[\Phi_i]=-\frac{1}{2}\int dt \int d^3x [B_i^a(\Phi_i)]^2
$$
in the Minkowski space for finite values of the momentum of the
rotator, so that the perturbation theory begins from
the zero value of the Minkowski action, in the contrast to the
instanton background.
This can be interpreted as the dominance of the monopole background.

For the self-dual BPS ansatz (\ref{bps})
the dominant value of the momentum of the rotator is
\be \label{dv}
\big|\frac{P_0}{2\pi}\big|=\frac{4\pi}{g^2}~.
\ee

\subsection{Hadronization}

In the lowest order of this perturbation theory, we can rewrite the
instantaneous interaction term in (\ref{wl2}) in the form of a
current-current interaction
\be \label{cc}
{\cal W}_{\rm int} =-\frac{1}{2}\int d^4x[D_k^{ab}({\Phi})\tilde{\sigma}^b]^2
=\frac{1}{2}\int d^4x \int d^3y
j_{\rm tot,0}^a(t,\vec{x})G^{ab}(\vec{x},\vec{y})j_{\rm tot,0}^b(t,\vec{y})~.
\ee
Following the QED perturbation theory of `radiative corrections' one
can formulate a similar perturbation theory of 'dynamical corrections'
in QCD. This 'dynamical perturbation theory' is based on the
decomposition of the  action $\tilde W$ for
the dynamical variables (\ref{ccg}) into the `free' part $\tilde W_0$,
the instantaneous interaction  ${\cal W}_{\rm{int}}$ (\ref{cc})
and the residual dynamical interaction $\tilde W_{\rm{int}}$
\be
\tilde W=\tilde W_0 +{\cal W}_{\rm{int}} +\tilde W_{\rm{int}}~.
\ee
The lowest order of the 'dynamical perturbation theory'
$\tilde W_{\rm{int}}=0$ describes  gluons and quarks in the monopole field
and their instantaneous bound states. Next orders of this perturbation
theory contain 'dynamical corrections' which describe matrix elements
of transitions between these states.

Bound states are obtained in a similar way as in QED
(i.e., by the Schwinger-Dyson and Bethe-Salpeter equations with
the instantaneous interaction $G^{bc}(\vec x,\vec y) )$
given in the Appendix.
In the field of the monopole the instantaneous quark-quark
potential is the sum of a Coulomb potential and a rising one.
It is well-known that the latter one leads to spontaneous chiral
symmetry breaking and to mesonic bound states.

\subsection{Frozen gluon approximation and U$_A(1)$- problem}

The `frozen gluon' approximation (FGA) follows from the Feynman path
integral if we neglect the dynamical gluon
fields: $\tilde A_i^a=0$ (\ref{ccg}).

The FGA generating functional (\ref{schb}) takes the form
\be\label{fga}
\tilde Z_{FGA}[\eta,\bar\eta]=\int \prod\limits_{t,x}\left\{
[d \bar\psi d \psi]\right\}
\exp \left\{i{\cal W} + i\tilde W_{FGA} +
i S[\eta, \bar \eta]\right\}~,
\ee
where
\be \label{wfga}
\tilde W_{FGA}=\tilde W_0+{\cal W}_{\rm{int}}
\ee
depends only on the quark fields.

The bilocal linearization of the four quark interaction ${\cal W}_{\rm{int}}$
using the Hubbard-Stratonovich transformation
%into two quark one and the integration over the quark fields
leads to an effective bilocal meson action \cite{had}.
This meson action includes Abelian anomalies
in the pseudoscalar isosinglet ($\eta_0$-meson) channel \cite{1,3,b}.

In our case, these anomalies
include the time derivative of topological variable
$N(t)$ due to interaction of the quark fields with the monopole
and the zero-mode ${\cal W}'$  (see eq. (\ref{gp})).

Neglecting all mesonic channels
except the $\eta_0$-meson one,
we get in FGA nothing but the well-known gauge-invariant expression
\be\label{ven}
\tilde W_{\rm anomaly}^{\eta_0}[\eta,\bar \eta]=C_{\eta}\int dt
\bar \eta(t,0) I_c\gamma_5\eta(t,0) \frac{g^2}{16 \pi^2}
\int d^3x G^a_{\mu \nu}{}^*G^a_{\mu \nu},
\ee
where $\eta,\bar \eta$ are fermion sources, $C_\eta$ is a constant.
In the BPS monopole selfdual field (\ref{class0}), (\ref{ass1}),
(\ref{bps})with
$G^a_{0i}=\dot N D_i^{ab}({\Phi})\Phi^b_0$, and
$2\pi \epsilon D_i^{ab}({\Phi})\Phi^b_0=B_i^a({\Phi})$
we obtain the normalizable zero-mode
\be \label{nor}
\frac{g^2}{16 \pi^2}
\int d^3x G^a_{\mu \nu}{}^*G^a_{\mu \nu}= \dot N(t)~,
\ee
as
$$
\frac{g^2}{8 \pi^2}\int d^3D_i^{ab}({\Phi})\Phi^b_0 B_i^a({\Phi})=1~.
$$
The physical anomaly term $\tilde W_{\rm anomaly}$ in the
$\eta_0$-meson channel  takes the form \cite{2}
\be\label{ven1}
\tilde W_{\rm anomaly}^{\eta_0}[\eta,\bar \eta]=
C_{\eta}\int dt \eta_0 \dot N,~~~
\eta_0(t)=\bar \eta(t) I_c\gamma_5\eta(t)~.
\ee
Following Veneziano \cite{2}, we can identify $\eta_0$ with the field of the
$\eta_0$-meson at rest. The effective action including the anomaly term
(\ref{ven}) and the zero-mode one (\ref{ktg}) is then
\be \label{eta}
W_{\rm eff}=\int dt [\frac{\dot N^2 {\cal I}}{2}+ \eta_0C_{\eta}\dot N +
\frac{\eta_0^2 m_0^2 V}{2}]~,
\ee
where $m_0$ is the standard current quark mass contribution to the $\eta_0$
meson mass. The diagonalization
of this effective action leads to an additional mass of $\eta_0$-meson
\be\label{u11}
\Delta m^2 V = \frac{C_{\eta}^2}{\cal I}~~~~\Rightarrow
~{\cal I}=\frac{C_{\eta}^2}{\Delta m^2 V}=
\frac{(2\pi)^2\epsilon}{g^2}4\pi.
\ee
The finite contribution of the zero-mode to the mass of the $\eta_0$-meson
entails the disappearance of the  mass of the collective topological
variable and leads to a stable perturbation theory in the infinite volume
limit, as in this limit the singularity-free BPS monopole converts into the
Wu-Yang one without singularities at the origin.

\subsection{Relativistic covariance}

Recall that Dirac has also obtained the unconstrained form of QED in terms of
gauge invariant variables for QED \cite{d} as functionals of the
initial gauge fields by explicitly resolving the Gauss law constraint.
The resulting unconstrained formulation of QED coincides with the one
obtained in the Coulomb gauge~\cite{hp} with the physical phenomena of
electrostatics, 'dressed' electrons, and two transverse photon degrees of
freedom. In QED, in terms of the Dirac variables~\cite{d}, the Poincare
symmetry is realized which is mixing with the gauge symmetry \cite{z}.
This mixing was interpreted in 1930 by Heisenberg and Pauli~\cite{hp}
(with reference to an unpublished note by von Neumann) as the transition
from the Coulomb gauge with respect to the time axis in the rest frame
($l_{\mu}^0=(1,0,0,0)$) to the Coulomb gauge with respect to the
time axis in the moving frame
$l_{\mu}=l_{\mu}^0 + \delta_L {l_{\mu}}^0 \;=\; {(L l^0)}_{\mu}$.
The Coulomb interaction has the covariant form
\begin{eqnarray}
{ W}_{C}
= \int d x d y \frac{1}{2}
j_{l}^{T}(x)
V_C(z^{\perp})
j_{l}^{T}(y) \delta(l \cdot z) \,\,\, ,
\end{eqnarray}
where
$
j_{l}^{T} = e \bar {\psi}^T \rlap/l \psi^T \,\, , \,\,
z_{\mu}^{\perp} =
z_{\mu} - l_{\mu}(z \cdot l) \,\, , \,\,
z_\mu = (x - y)_\mu  \,\, . \,\,
$
This transformation law and the relativistic covariance of this formulation
of QED has been proven by Zumino \cite{z} on the level of the algebra of
generators of the Poincare group.

In QED we can change variables to construct the generating functional
of the Green functions in any gauges including the Lorentz invariant ones.
The invariance of the corresponding Green functions under a change of
variables (which generates the Ward-Taylor-Slavnov identities \cite{fs})
is  garanteed by the Dirac factors in source terms, which
restore the Coulomb gauge Feynman rules in any Lorentz invariant gauge.
So, the Coulomb interaction and electrostatics are consequences
of the identification of the physical degrees of freedom which correspond
to an explicit solution of the Gauss law, but not primarily to the choice of
the gauge. For example,
if one would omit the Dirac factors in the source terms in relativistically
invariant Lorentz gauge formulations of QED, one would get
the Wick-Cutkosky bound states formed by gauge propagators with
light-cone singularities with a spectrum different
\footnote{One of the authors (V.P.) thanks W. Kummer who pointed out that in
Ref. \cite{kum} the difference between the Coulomb atom and
the Wick-Cutkosky bound states in QED has been demonstrated.}
from the observed one which corresponds to the instantaneous Coulomb
interaction.

In QCD, the moving Lorentz frame corresponds to the moving
Wu-Yang monopole \cite{fn}.
We should only manage to realize this transformation law on the level
of operators.

\section{Conclusion}

In this paper, we have shown that there is a
collective topological excitation in the gluon spectrum as the zero-mode
of Gauss' constraint from the winding number class of functions of
the large gauge transformations.

This topological excitation leads to the dominance of the Wu-Yang monopole
in the Feynman path integral compatible with the
equivalent unconstrained system obtained in this work.

In the field of the Wu-Yang monopole the instantaneous quark-quark
potential is the sum of a Coulomb-type potential and a rising one.
The latter one leads to spontaneous chiral symmetry breaking and to
mesonic bound states.

The $\eta_0$-meson mixes with the zero-mode so that after
diagonalization of this low-energy action a mass shift of the
$\eta_0$-meson is obtained which resolves the $U_A(1)$ problem.

Colored amplitudes contain additional phase factors
which depend on the zero-mode.
Averaging  the colored amplitudes over the zero-mode parameters
can lead to the phenomenon of complete destructive interference
\cite{vp1,vp3}, so that the color amplitudes disappear.
The colorless ones of the type
of the expectation values of the electroweak currents do not vanish
since the zero-mode  phase factors are absent.

According to Heisenberg, Pauli \cite{hp} and Zumino \cite{z},
the relativistic covariance is established by a rotation of the timelike
axis so that the non-Abelian Coulomb-gauge field moves together with the
relativistic bound states.
Recently, Faddeev and Niemi~\cite{fn} constructed a similar
relativistically covariant form of the Wu-Yang monopole.

In summary, the present scheme for the introduction of topological
global  variables is a promising tool for the investigation
of the challenging properties of QCD such as the meson spectrum,
chiral symmetry breaking, quark and gluon confinement, and the $U_A(1)$
anomaly.

%\newpage
\section*{Acknowledgements}

\medskip

We thank  Profs. A. V. Efremov, G. A. Gogilidze, N. Ilieva, V. G.
Kadyshevsky, A. M. Khvedelidze, E. A. Kuraev, M. Lavelle, D. McMullan and
W. Thirring for interesting and critical discussions.
One of us (VNP) is grateful for a stipend from the
{Max-Planck-Gesellschaft} for his study visit at the University of
Rostock.

\bigskip

\begin{appendix}
\section{The Green function}

\medskip
%++++++++++++++++++++++++++++++++++++++++%
%\subsection{Non-Abelian generalization of the Coulomb potential}
%++++++++++++++++++++++++++++++++++++++++%
We can calculate the instantaneous
the Green function (\ref{gaub01})
 \begin{equation}\label{gaub01}
(D^2(b))^{ab}({\vec x})G^{bc}(\vec x,\vec y)=\delta^{ac}\delta^3(x-y)~.
\end{equation}
In the presence of the Wu-Yang monopole we have
$$
(D^2)^{ab}({\vec x})= \delta^{ab}\Delta-
\frac{n^an^b+\delta^{ab}}{r^2}+
2(\frac{n_a}{r}\partial_b-\frac{n_b}{r}\partial_a)~,
$$
and $n_a(x)={x_a}/{r};~ r=|{\vec x}|~$.
Let us decompose $G^{ab}$ into a complete set of orthogonal vectors in
color space
$$
G^{ab}({\vec x},{\vec y})=  [n^a(x)n^b(y)V_0(z) +
\sum\limits_{\alpha=1,2}
e^a_{\alpha}(x) e^b_{\alpha}(y)V_1(z)];~~~(z=|{\vec x}-{\vec y}|)~.
$$
Substituting the latter into the first equation, we get
$$
\frac{d^2}{dz^2}V_n+ \frac{2}{z} \frac{d}{dz}V_n- \frac{n}{z^2}V_n=0
~~~n = 0,~1~.
$$
The general solution for the last equation is
\begin{equation} \label{pot}
 V_n(z)= d_n z^{l_1^n} + c_n z^{l_2^n}~,
\end{equation}
where $d_n$, $c_n$ are constants, and
${l_1^n},~{l_2^n}$ can be found as  roots of the equation
$(l^n)^2 + l^n =n$~, i.e.
\begin{equation} \label{l1}
{l_1^n} =-\frac{1+\sqrt{1+4n}}{2};~~~
{l_2^n} =\frac{-1+\sqrt{1+4n}}{2}~.
\end{equation}
It is easy to see that for $n=0$ we get the Coulomb-type potential
$d_0=-1/4\pi$,
and for $n=1$  the 'golden section' potential with
\begin{equation} \label{fun}
{l_1^1} =-\frac{1+\sqrt{5}}{2}\approx - 1.618~
 ;~~~{l_2^1} =\frac{-1+\sqrt{5}}{2}\approx 0.618~.
\end{equation}
The last potential (in the contrast with the Coulomb-type one)
leads to  spontateous chiral symmetry breaking and can be used
 as the potential for the 'hadronization' of quarks and gluons
in QCD \cite{vp3,had}.
\end{appendix}

\newpage
%\Bibliography{99}

\end{document}